\documentclass{article}
\usepackage{float}
\usepackage{hyperref}
\usepackage{algorithm}
\usepackage{algpseudocode}
\usepackage{arxiv}
\usepackage{amsmath}
\usepackage[utf8]{inputenc} 
\usepackage[T1]{fontenc}    
\usepackage{hyperref}       
\usepackage{url}            
\usepackage{booktabs}       
\usepackage{amsfonts}       
\usepackage{nicefrac}       
\usepackage{microtype}      
\usepackage{lipsum}
\usepackage{graphicx}
\graphicspath{ {./images/} }
\title{Generalized first order rational ODE reduction algorithm with bounded degree transformation}

\author{
 Shaoxuan Huang \\
  Chinese university of Hongkong, ShenZhen
  \texttt{223010044@link.cuhk.edu.cn} \\
}
\begin{document}
\maketitle
\begin{abstract}
As a generalization of our previous result\cite{huang2025algorithm}, this paper aims to answer the following question: Given a 2-dimensional polynomial vector field $y^{\prime}=\frac{M(x,y)}{N(x,y)}$, how to find a rational transformation $y \to \frac{A(x,y)}{B(x,y)}$ with bounded degree numerator, the inverse of which transforms this vector field into a simpler form $y^{\prime}=\sum_{i=0}^nf_i(x)y^i$.  Such a structure, often known as the generalized Abel equation and has been studied in various areas, provides a deeper insight into the property of the original vector field.
We have implemented an algorithm with considerable performance to tackle this problem and the code is available in \href{https://www.researchgate.net/publication/393362858_Generalized_ODE_reduction_algorithm}{Researchgate}.
\end{abstract}
\section{Review of the existing method}
The rational differential equation, which can be viewed as a planar autonomous system, takes the form:
\begin{equation}
    \frac{dy}{dx} = \frac{M(x,y)}{N(x,y)},
\end{equation}
and remains an important subject of research in dynamical systems and physics. For methods to compute the first integral of first-order ODEs, see \cite{goriely2001integrability, kamke2013differentialgleichungen} for an overview.

A special case of (1) that has attracted particular interest is:
\begin{equation}
    y^{\prime} = \sum_{i=0}^n f_i(x) y^i,
\end{equation}
where \( f_i(x) \) are rational functions in \( x \). This class includes Bernoulli, Chini, Riccati, and Abel ODEs as special cases. Its integrability can be characterized by equivalence classes under linear transformations; see \cite{appell1889invariants, CHEBTERRAB2000204, cheb2000first} for details:
\begin{equation}
    \left\{x \to F(x),\, y \to P(x)y + Q(x)\right\}.
\end{equation}

Such ODEs are often studied in broader contexts, such as the existence of Darboux polynomials or limit cycles \cite{demina2020classifying}. In \cite{huang2025algorithm}, we consider a more general transformation applied to (2):
\begin{equation}
    \left\{x \to F(x),\, y \to \frac{A(x,y)}{B(x,y)}\right\}.
\end{equation}
Applying this transformation yields an ODE with a more complex structure:
\begin{equation}
    y^{\prime} = \frac{\sum_{i=0}^{n} f_i A^{i} B^{n-i} - B^{n-2}\left(B \frac{\partial A}{\partial x} - A \frac{\partial B}{\partial x}\right)}{B^{n-2}\left(B \frac{\partial A}{\partial y} - A \frac{\partial B}{\partial y}\right)}.
\end{equation}

In \cite{huang2025algorithm}, we assume the rational ODE (1) is exactly of the form (5), with coprime numerator and denominator. We develop new methods to detect transformations \( y \to \frac{A(x,y)}{B(x,y)} \); if such a transformation exists, the ODE reduces to (2). However, if \( A(x,y) \) has an irreducible factor \( p_i(x,y) \) with multiplicity \( \alpha_i > 1 \) and \( f_0(x) = 0 \), then \( p_i(x,y)^{\alpha_i-1} \) becomes a common factor of the numerator and denominator. Canceling this factor disrupts the structure (5), causing the original algorithm to fail.

In this paper, we address the following problem: Given \( M \) and \( N \) in a rational ODE and a degree bound for \( A(x,y) \), we aim to determine whether there exist polynomials \( A(x,y) \), \( B(x,y) \), \( f_i(x) \), \( t(x) \), and \( c(x,y) \) (with \( A \) and \( B \) coprime in \( \mathbb{R}[x,y] \) and \( c(x,y) \mid A(x,y) \)) such that:
\begin{equation}
    \begin{aligned}
        M c &= \sum_{i=0}^n f_i A^i B^{n-i} - t \left(B \frac{\partial A}{\partial x} - A \frac{\partial B}{\partial x}\right) B^{n-2}, \\
        N c &= t \left(B \frac{\partial A}{\partial y} - A \frac{\partial B}{\partial y}\right) B^{n-2}.
    \end{aligned}
\end{equation}

Note that when \( f_0(x) = 0 \), \( A(x,y) \) is a Darboux polynomial of the ODE \( y^{\prime} = \frac{M(x,y)}{N(x,y)} \). The degree bound for Darboux polynomials or first integrals is generally difficult to determine. A key question is whether an efficient algorithm exists to compute them given a tight degree bound. Existing methods based on Gröbner bases or regular chains \cite{cheze2020symbolic, duarte2021efficient} suffer from rapidly growing computational costs as the ODE size increases. Under assumption (6), we demonstrate that \( A(x,y) \) and \( B(x,y) \) can be computed via linear equations, reducing the ODE to (2). This simplification makes determining first integrals or Darboux polynomials significantly easier than for the general form (1).
\section{New algorithm considering canceled factor}
\subsection{description of the new method}
The key idea of our algorithm is to treat the canceled factor and $A$ separately, even though the canceled factor is itself a factor of $A$. Since it divides $A$, its degree in $y$ must be strictly less than that of $A$ in $y$. This allows us to search for the degree of the canceled factor in $y$ from $0$ up to $\deg_y(A) - 1$. For each value in this range, we augment the $y$-degrees of $M$ and $N$ to reconstruct their original form (5) before cancellation.  \\
Using these adjusted degrees, we apply the degree test procedure from \cite{huang2025algorithm} to determine possible values for $n$ and candidate forms of $B$. For each candidate, we solve the following equation for the undetermined coefficients in $c(x,y)$ and $A(x,y)$:
\begin{equation}
    Nc = t\left(B\frac{\partial A}{\partial y} - A\frac{\partial B}{\partial y}\right)B^{n-2},
\end{equation}
which reduces to a linear system. In certain cases, solving (7) suffices to determine both $A$ and $c$. \\
\textbf{Example 2.1}
Consider the Abel differential equation (class 4, solvable via Liouvillian functions):
\begin{equation}
    y^{\prime} = y^3 - \frac{x+1}{x}y^2.
\end{equation}
Applying the transformation
\begin{equation}
    \left\{y \mapsto \frac{(y+x+1)^2(y^2+x-1)}{(xy-2)^2(y+x^2-1)^2},\, x \mapsto x\right\}
\end{equation}
yields an ODE $y^{\prime} = \frac{M(x,y)}{N(x,y)}$ with $\deg(M) = \deg(N) = 19$, where:
\begin{align*}
    M &= 4x^{13}y^6 + 6x^{12}y^7 + \cdots + 7x - 4y - 4, \\
    N &= 2x\left(x^{12}y^6 - x^{14}y^3 + \cdots + 8x + 112y - 32\right).
\end{align*}
Assuming $\deg(A) = 4$, our algorithm successfully identifies candidates for the canceled factor and $B$ when $n=3$:
\begin{align*}
    c &= x^3y b_{3,1} + x^3 b_{3,0} + x^2y b_{2,1} + \cdots + y b_{0,1} + b_{0,0}, \\
    B &= (xy-2)^2(y+x^2-1)^2.
\end{align*}
With $A$ expressed as:
\begin{equation*}
    A = \sum_{\substack{0\leq i+j \leq 4}} a_{i,j}x^i y^j,
\end{equation*}
substitution into the condition
\begin{equation}
    Nc = x\left(B\frac{\partial A}{\partial y} - A\frac{\partial B}{\partial y}\right)B
\end{equation}
yields the solution:
\begin{align*}
    A &= b_{1,0}(x^2y^2 + 2xy^3 + y^4 + x^3 + 2x^2y + 3xy^2 + 2y^3 + x^2 - x - 2y - 1), \\
    c &= b_{1,0}(x + y + 1).
\end{align*}
Here, $b_{1,0}$ is an arbitrary nonzero constant. Choosing $b_{1,0}=1$ and substituting into
\begin{equation}
    Mc = \sum_{i=0}^n f_i A^i B^{n-i} - x\left(B\frac{\partial A}{\partial x} - A\frac{\partial B}{\partial x}\right)B,
\end{equation}
we recover the reduced ODE:
\begin{equation}
    y^{\prime}x = y^3x - (x+1)y^2.
\end{equation}
\subsection{cases that requires further reduction of parameters}
In general, solving (7) may not suffice to determine $A$ and $c$ completely, as it does not enforce the divisibility condition $c \mid A$. When the canceled factor $c$ has high degree and constitutes a significant portion of $A$, the solution may remain overly general, requiring further reduction. To address this, we compute the remainder of $A$ divided by $c$ using either the $\mathrm{tdeg}(y,x)$ or $\mathrm{tdeg}(x,y)$ monomial order. \\
\textbf{Example 2.2} Consider the Abel differential equation from Example 2.1, but now applying the transformation:
\begin{equation}
    \left\{y \to \frac{(y+x+1)^4}{(xy-2)^3(y+x^2-1)},\, x \to x \right\}
\end{equation}
This yields an ODE $y^{\prime} = \frac{M(x,y)}{N(x,y)}$ with $\deg(M) = \deg(N) = 17$, where:
\begin{align*}
    M &= x^{11}y^6 + 5x^{10}y^7 + \cdots - 303x - 32y - 8, \\
    N &= x\left(-3x^{11}y^5 + x^{10}y^6 - \cdots + 160x - 512y + 320\right).
\end{align*}
The canceled factor has degree 3. Setting
\begin{align*}
    A &= \sum_{\substack{0\leq i+j \leq 4}} a_{i,j}x^i y^j, \\
    c &= \sum_{\substack{0\leq i+j \leq 3}} b_{i,j}x^i y^j,
\end{align*}
and substituting the candidate $B = (xy-2)^3(y+x^2-1)$ into
\begin{equation}
    Nc = t\left(B\frac{\partial A}{\partial y} - A\frac{\partial B}{\partial y}\right)B,
\end{equation}
we obtain the parametric solution:
\begin{align*}
    A &= b_{3,0}x^4 + \left(\frac{3}{2}b_{2,1} - \frac{1}{2}b_{3,0}\right)x^3y + \cdots + \left(\frac{1}{3}b_{2,1} + 3b_{3,0}\right)y + \frac{4}{3}b_{2,1} - 3b_{3,0}, \\
    c &= b_{3,0}x^3 + b_{2,1}x^2y + b_{2,1}xy^2 + \frac{1}{3}b_{2,1}y^3 + \left(\frac{5}{6}b_{2,1} + \frac{1}{2}b_{3,0}\right)x^2 \\
    &\quad + \left(\frac{3}{2}b_{2,1} + \frac{3}{2}b_{3,0}\right)xy + b_{2,1}y^2 + \left(\frac{7}{6}b_{2,1} - \frac{1}{2}b_{3,0}\right)x + b_{2,1}y + \frac{1}{3}b_{2,1}.
\end{align*}
The parameters $b_{3,0}$ and $b_{2,1}$ require further selection to ensure $c \mid A$ and to solve the reduced ODE. Computing the remainder of $A$ divided by $c$ and setting it to zero yields a linear relationship between $b_{3,0}$ and $b_{2,1}$, leading to a valid solution. \\
Note that the division may fail in symbolic computation systems like Maple, which implicitly assumes all parameters to be non-zero. To obtain useful results, we must:
\begin{itemize}
    \item Identify all combinations of non-zero parameters in $A$ and $c$
    \item Set other variables to zero when appropriate
    \item Discard trivial solutions where all parameters vanish
\end{itemize}
\subsection{cases where all factors in $A$ have same multiplicity}
An interesting special case occurs when the numerator polynomial $A$ can be expressed as $\prod_i p_i(x,y)^k$, where all irreducible factors share the same multiplicity $k$. In this scenario, we have:
$$
    A = P^k \quad \text{and} \quad c = P^{k-1},
$$
where $P$ is the product of distinct irreducible factors of $A$. Under this assumption, the polynomial $N$ takes the form:
\begin{equation}
    N = \left(kB\frac{\partial P}{\partial y} - P\frac{\partial B}{\partial y}\right)B^{n-2}.
\end{equation}
Substituting valid candidates for $B$ into (15) allows us to solve for $P$ directly through a linear system. We refer to this as the \textbf{power case solution}. To identify such cases algorithmically, we verify whether $\deg_y(A)$ is divisible by $\deg_y(A) - \deg_y(c)$.
\subsection{Selection of $t$}
In our previous analysis, we assumed $t$ to be the product of all $x$-dependent factors. However, this assumption may not hold when there exists $a \in \mathbb{C}$ such that $A(a,y) = B(a,y)$. In such cases, the expression $B\frac{\partial A}{\partial y} - A\frac{\partial B}{\partial y}$ will contain $(x-a)$ as a factor, which should be excluded from $t$. We observe the following implication: $A(a,y) = B(a,y) \implies B(a,y) \mid M(a,y).$ \\
This observation leads to an improved candidate selection procedure:
\begin{algorithmic}[1]
    \State For each factor $(x-a)^k$ in $t$ (where $k$ is its multiplicity)
    \State Verify if $B(a,y) \mid M(a,y)$
    \If{condition holds}
        \State Add $\frac{t}{(x-a)^m}$ to candidate list for all $1 \leq m \leq k$
    \EndIf
\end{algorithmic}
This refinement ensures we properly account for the special case where $A$ and $B$ coincide at certain points $x = a$, leading to more accurate candidate selection for $t$.
To summarize, we give a pseudo code for reduction algorithm for $n > 2$. \\
\begin{scriptsize}
\begin{algorithm}[H]
    \caption{reduction algorithm for ODE $y^{\prime}=\frac{M(x,y)}{N(x,y)}$}
    \begin{algorithmic}[1]
        \Statex \textbf{Input:} Polynomial $M(x,y)$ and $N(x,y)$ in a rational ODE $y^{\prime}=\frac{M(x,y)}{N(x,y)}$ and the degree bound degreeA for A(x,y)
        \Statex \textbf{Output:} A reduced ODE $y^{\prime}t(x)=\sum_{i=0}^nf_i(x)y^i$ and the reduction transformation $y \to \frac{A(x,y)}{B(x,y)}$ when n>2
        \Statex i := degree(M,y)
        \Statex j := degree(N,y)
        \Statex t(x) := product of all factors of $N$ that depend on $x$ only
        \Statex factorList := all factors that depend on $y$
        \Statex multList := multiplicities of all factors that depend on $y$
        \Statex Construct candidates of A(x,y) with undetermined coefficients
        \Statex \textbf{for} cany \textbf{from} 0 \textbf{to} degreeA-1 \textbf{do:}
        \Statex \quad inew := i+cany;
        \Statex \quad jnew := j+cany;
        \Statex \quad Construct candidates of canceled factor of degree cany in y
        \Statex \quad \textbf{for} index1 \textbf{from} 3 \textbf{to} $\max(\text{multList})+2$ \textbf{do:}
        \Statex \quad \quad \textbf{for} index2 \textbf{from} $\max(\frac{\text{inew}}{\text{index1}},\frac{\text{jnew+1}}{\text{index1}})$ \textbf{to} jnew \textbf{do:}
        \Statex \quad \quad \quad \textbf{for} index 3 \textbf{from} 0 \textbf{to} index2-1 \textbf{do:}
        \Statex \quad \quad \quad \quad \textbf{if} index2+index3-1+(index1-2)index3=jnew \textbf{and} inew = index1 $\cdot$ index2 \textbf{and} index2 > cany \textbf{and} index2 $\leq$ degreeA \textbf{then:}
        \Statex \quad \quad \quad \quad \quad Choose possible factors of B
        \Statex \quad \quad \quad \quad \quad Construct candidates of B of degree index3
        \Statex \quad \quad \quad \quad \quad Check for power case solution when index2 divide by index2 - cany
        \Statex \quad \quad \quad \quad \quad for each candidate of B construct candidates of t
        \Statex \quad \quad \quad \quad \quad Verifying candidates of B,t and solve for A and canceled factor
        \Statex \quad \quad \quad \quad \quad use B,A,canceled factor to solve for $f_i(x)$
        \Statex \quad \quad \quad \quad \textbf{end if}
        \Statex \quad \quad \quad \quad \textbf{if} index2+index3-1+(index1-2)index2=jnew \textbf{and} (inew = index1 $\cdot$ index2 \textbf{or} inew $\leq$ jnew+1) \textbf{and} index2 > cany \textbf{and} index3 $\leq$ degreeA \textbf{then:}
        \Statex \quad \quad \quad \quad \quad Choose possible factors of B
        \Statex \quad \quad \quad \quad \quad Construct candidates of B of degree index2
        \Statex \quad \quad \quad \quad \quad Check for power case solution when index2 divide by index2 - cany or index3 divide by index3 - cany
        \Statex \quad \quad \quad \quad \quad for each candidate of B construct candidates of t
        \Statex \quad \quad \quad \quad \quad Verifying candidates of B,t and solve for A and canceled factor 
        \Statex \quad \quad \quad \quad \quad use B,A,canceled factor to solve for $f_i(x)$
        \Statex \quad \quad \quad \quad \textbf{end if}
        \Statex \quad \quad \quad \textbf{end do}
        \Statex \quad \quad \textbf{end do}
        \Statex \quad \textbf{end do}
        \Statex \textbf{end do}
    \end{algorithmic}
\end{algorithm}
\end{scriptsize}
\subsection{Computing inverse integrating factor when $n=2$}
When $n=2$, $B$ typically does not divide $N$, preventing the use of degree test procedures to determine $B$. However, in this case, the reduced ODE takes one of two classical forms:
\begin{itemize}
    \item When $f_0 = 0$, the reduced ODE becomes a Bernoulli equation, admitting a Liouvillian inverse integrating factor of the form \cite{duarte2021efficient}:
    \begin{equation}
        \mu(x,y) = e^{f(x)}\prod_i p_i(x,y)^{\alpha_i}, \quad \alpha_i \in \mathbb{N}.
    \end{equation}
    \item When $f_0 \neq 0$, no factors are canceled and the reduced ODE remains a Riccati equation.
\end{itemize}
For the Bernoulli case ($f_0=0$), \cite{duarte2021efficient} developed a semi-algorithm to find inverse integrating factors. The approach involves:
\begin{enumerate}
    \item Constructing an auxiliary ODE $y' = \frac{M_1(x,y)}{N_1(x,y)}$ with first integral $\mu(x,y)$ and the inverse integrating factor $\tau(x,y)$
    \item Proving $\tau(x,y)$ is polynomial in $x$ and $y$
    \item Establishing the fundamental equation:
    \begin{equation}
        \Delta - T_0\tau = 0,
    \end{equation}
    where
    \begin{align*}
        \Delta &= MN_1 - M_1N, \\
        T_0 &= \frac{\partial M}{\partial y} + \frac{\partial N}{\partial x}.
    \end{align*}
\end{enumerate}
Given degree bounds on $M_1$, $N_1$, and $\tau$, we solve the system comprising (17) and the inverse integrating factor condition:
\begin{equation}
    M_1\frac{\partial \tau}{\partial y} + N_1\frac{\partial \tau}{\partial x} = \tau\left(\frac{\partial M_1}{\partial y} + \frac{\partial N_1}{\partial x}\right).
\end{equation}
For the general Riccati case ($f_0 \neq 0$), \cite{huang2025algorithm} presents a complete algorithm for computing inverse integrating factors without requiring degree bounds.
\section{Code implementation and Sample output}
We have implemented this algorithm in Maple and the code implementation is available in \href{https://www.researchgate.net/publication/393362858_Generalized_ODE_reduction_algorithm}{Researchgate}. To use it, read it from the txt files to import all the modules. Here we display some sample input/output of this program: \\
\textbf{Example 3.1} 
Consider the following transformation:
$$
\left\{y \to \frac{\left(x \,y^{2}+\left(x +1\right) y +3 x^{2}+x -1\right)^{3}}{\left(y x +x -1\right)^{2} \left(y +x +1\right)^{2}}, x \to x \right\}
$$
applied to the ODE:
$$
y^{\prime}x=y^4-(x^2+1)y^2
$$
For the new ODE $y^{\prime}=\frac{M(x,y)}{N(x,y)}$, execute the command \texttt{degree-test(M,N,9)}, we compute the transformation in 0,266 seconds: \\
\texttt{
canceled factor's degree in y is 0 \\
the value of n is 4 \\
the degree of A in y is 5 \\
the degree of B in y is 3 \\
Solving for A \\
Solutions for A,canceled factor B and t found are: \\
$$
                               []
$$
canceled factor's degree in y is 1 \\
the value of n is 3 \\
the degree of A in y is 1 \\
the degree of B in y is 7 \\
Solving for A \\
Solutions for A,canceled factor B and t found are: \\
$$
                               []
$$
the value of n is 3 \\
the degree of A in y is 7 \\
the degree of B in y is 4 \\
Solving for A \\
Solutions for A,canceled factor B and t found are: \\
$$
                               []
$$
the value of n is 7 \\
the degree of A in y is 3 \\
the degree of B in y is 2 \\
Solving for A \\
Solutions for A,canceled factor B and t found are: \\
$$
                               []
$$
canceled factor's degree in y is 2 \\
canceled factor's degree in y is 3 \\
canceled factor's degree in y is 4 \\
the value of n is 3 \\
the degree of A in y is 2 \\
the degree of B in y is 8 \\
Try to find power case solutions... \\
Solving for A \\
Solutions for A,canceled factor B and t found are: \\
$$
                               []
$$
the value of n is 3 \\
the degree of A in y is 8 \\
the degree of B in y is 5 \\
Try to find power case solutions... \\
Solving for A \\
Solutions for A,canceled factor B and t found are: \\
$$
                               []
$$
the value of n is 4 \\
the degree of A in y is 0 \\
the degree of B in y is 6 \\
Try to find power case solutions... \\
Solving for A \\
Solutions for A,canceled factor B and t found are: \\
$$
                               []
$$
the value of n is 4 \\
the degree of A in y is 6 \\
the degree of B in y is 4 \\
Try to find power case solutions... \\
Power case successful. The reduced ODE is:
$$
x \left(\frac{d}{d x}y \! \left(x \right)\right) = 
\left(-x^{2}-1\right) y \! \left(x \right)^{2}+y \! \left(x \right)^{4}
$$
$$
\left[4, \left[
\frac{\left(x \,y^{2}+\left(x +1\right) y +3 x^{2}+x -1\right)^{3}}{\left(y x +x -1\right)^{2} \left(y +x +1\right)^{2}}
, x \left(\frac{d}{d x}y \! \left(x \right)\right) = 
\left(-x^{2}-1\right) y \! \left(x \right)^{2}+y \! \left(x \right)^{4}
\right]\right]
$$
}
\\
\textbf{Example 3.2} Consider the following transformation:
$$
 \left\{ y \to \frac{\left(y +x +1\right)^{3} \left(y x +2\right)^{2}}{\left(y x +3 x -1\right) \left(x^{2}+y +1\right)^{4}}, x \to x \right\}
$$
applied to the ODE:
$$
y^{\prime}x=y^5-(x^2+1)y^2
$$
The numerator and denominator of the transformation coincide at the point $x=1$, therefore produces an "unexpected" factor $x-1$ in $N(x,y)$ of the new ODE $y^{\prime}=\frac{M(x,y)}{N(x,y)}$. Execute the command \texttt{degree-test(M,N,7)}, we compute the transformation, correct $t$ and the reduced ODE in 4.75 second. \\
\texttt{
canceled factor's degree in y is 0 \\
the value of n is 11 \\
the degree of A in y is 1 \\
the degree of B in y is 2 \\
Solving for A \\
Solutions for A,canceled factor B and t found are: \\
$$
                               []
$$
canceled factor's degree in y is 1    \\
canceled factor's degree in y is 2    \\
the value of n is 3    \\
the degree of A in y is 7     \\
the degree of B in y is 8    \\
Solving for A   \\
Solutions for A,canceled factor B and t found are:
$$
                               []
$$
the value of n is 4 \\
the degree of A in y is 5 \\
the degree of B in y is 6 \\
Solving for A \\
Solutions for A,canceled factor B and t found are: \\
$$
                               []
$$
the value of n is 6 \\
the degree of A in y is 3 \\
the degree of B in y is 4 \\
Try to find power case solutions... \\
Try to find power case solutions... \\
Solving for A \\
Solutions for A,canceled factor B and t found are: 
$$
                               []
$$
the value of n is 8 \\
the degree of A in y is 2 \\
the degree of B in y is 3 \\
Try to find power case solutions... \\
Solving for A \\
Solutions for A,canceled factor B and t found are:
$$
                               []
$$
canceled factor's degree in y is 3 \\
the value of n is 5 \\
the degree of A in y is 4 \\
the degree of B in y is 5 \\
Try to find power case solutions... \\
Solving for A \\
Try to reduce parameters in candidates of A and canceled factors... \\
Try to reduce parameters in candidates of A and canceled factors... \\
Successfully reduce parameters,the reduced candidates are
$$
\left[
\left(y +x +1\right)^{3} \left(y x +2\right)^{2},\left(y +x +1\right)^{2} \left(y x +2\right)
, \left(x^{2}+y +1\right)^{4} \left(y x +3 x -1\right)\right]
$$
Try to reduce parameters in candidates of A and canceled factors... \\
Solutions for A,canceled factor B and t found are:
$$
\left[\left[
\left(y +x +1\right)^{3} \left(y x +2\right)^{2},\left(y +x +1\right)^{2} \left(y x +2\right)
, \left(x^{2}+y +1\right)^{4} \left(y x +3 x -1\right), x\right]\right
]
$$
Reduction transformation successful. The reduced ODE is:
$$
x \left(\frac{d}{d x}y \! \left(x \right)\right) = 
\left(-2 x^{2}-2\right) y \! \left(x \right)^{2}+16 y \! \left(x \right)^{5}
$$
$$
\left[5, \left[
\frac{\left(y +x +1\right)^{3} \left(y x +2\right)^{2}}{\left(x^{2}+y +1\right)^{4} \left(y x +3 x -1\right)}
, x \left(\frac{d}{d x}y \! \left(x \right)\right) = 
\left(-2 x^{2}-2\right) y \! \left(x \right)^{2}+16 y \! \left(x \right)^{5}
\right]\right]
$$
} \\
\textbf{Example 3.3.} Consider the following transformation:
$$
 \left\{ y \to \frac{\left(y +x +1\right)^{3} \left(y x +2\right)^{2}}{\left(y x +3 x -1\right) \left(x^{2}+y +1\right)^{4}}, x \to x \right\}
$$
applied to the ODE:
$$
y^{\prime}=y^2-(x^2+1)y
$$
Using \texttt{degree-test(M,N,7)}, our algorithm computes a Liouvillian inverse integrating factor in 0.312 seconds:
\texttt{
canceled factor's degree in y is 0 \\
canceled factor's degree in y is 1 \\
the value of n is 4 \\
the degree of A in y is 1 \\
the degree of B in y is 2 \\
Try to find power case solutions... \\
Solving for A \\
Solutions for A,canceled factor B and t found are: 
$$
                               []
$$
canceled factor's degree in y is 2 \\
the value of n is 3 \\
the degree of A in y is 2 \\
the degree of B in y is 3 \\
Try to find power case solutions... \\
Solving for A \\
Solutions for A,canceled factor B and t found are:
$$
                               []
$$
canceled factor's degree in y is 3 \\
canceled factor's degree in y is 4 \\
canceled factor's degree in y is 5 \\
canceled factor's degree in y is 6 \\
Computing inverse integrating factor when n = 2: handling cases factors being canceled... \\
Inverse integrating factor n=2 successful
$$
\left[2, 
\left(y x +2\right)^{3} \left(y +x +1\right)^{4} {\mathrm e}^{\frac{x \left(x^{2}+3\right)}{3}}
\right]
$$
}
\section{Concluding remark and future work}
The current work suggests several important directions for future research. One promising avenue involves developing algorithms that rely solely on the degree of $y$ in $A$ to solve for the coefficients $a_i(x)$, $c_i(x)$, and $f_i(x)$ from Equations (6). While computationally more intensive than solving for constant coefficients, this approach would prove valuable when information about $\deg_x(A)$ is unavailable. Such development could significantly expand the applicability of our methods to cases with more complex coefficient structures.

Another critical direction concerns transformations with $y$-independent denominators $B(x)$. The problem becomes particularly challenging when $M$ and $N$ share common factors, requiring the simultaneous solution of both $M$ and $N$ equations through computationally expensive calculations for solving coefficients in two remainders.

Our current assumption that canceled factors must divide the numerator $A$ generally holds, but notable exceptions exist. For instance, in the equation
$ \frac{dy}{dx} = \frac{(y^2+1)(y^2-2y-1)}{-4yx^3+2y^2x+4xy-2y^2-2x-4y+2}$,
 $y^2+1$ is canceled without it being a numerator factor of the minimum degree transformation. Characterizing such exceptions presents both theoretical interest and practical importance, potentially leading to more robust algorithms that can handle these special cases while maintaining computational efficiency.
\bibliographystyle{plain}  
\bibliography{references}
\end{document}